\documentclass[amsmath,amssymb]{nature}

\usepackage{ifthen}		
\newboolean{SubmittedVersion}
\setboolean{SubmittedVersion}{False} 	

\usepackage{graphicx}
\usepackage{amsmath}
\usepackage{amssymb}
\DeclareMathAlphabet{\mathpzc}{OT1}{pzc}{m}{it}

\title{Simulating topological phases and topological phase transitions with classical strings}
\author{Yi-Dong Wu$^1$}

\begin{document}

\maketitle

\begin{affiliations}
 \item Department of Applied Physics, Yanshan University, Qinhuangdao, Hebei, 066004, China
 \end{affiliations}

\begin{abstract}
The discovery of the topological insulators has fueled a surge of interests in the topological phases in periodic systems. Topological insulators have bulk energy gap and topologically protected gapless edge states\cite{Kane2005a,Kane2005b,Konig2007}. The edge states in electronic systems have been detected by observing the transport properties the Hgte quantum wells\cite{Konig2007,Roth2009}. The electromagnetic analogues of such electronic edge states have been predicted and observed in photonic crystals, coupled resonators and linear circuits\cite{Haldane,Wang2009,Hafezi2013,Rechtsman,Mei,Khanikaev2013,Liang2013,
Albert,Ningyuan}. However, the edge state spectrums of the two dimensional insulators and their electromagnetic analogues haven't been directly measured and thus the gaplessness of the edge states hasn't been experimentally confirmed. Here I show the classical strings are more convenient choice to study the topological phases than the electromagnetic waves. I found that classical strings with periodic densities can simulate a variety of topological phases in condensed matter physics including two dimensional topological insulators\cite{Kane2005a,Kane2005b,Konig2007}, three dimensional topological semimetals\cite{Wang,Liu,Wan,Burkov2011} and weak topological insulators\cite{Fu}. Because the eigenfrequencies and eigenfunctions of the strings can be easily measured, not only the gapless edge state spectrum but also the bulk topological invariant can now be directly observed. Further more I show the topological phase transitions can be simulated by the strings. My results show the richness of topological phases of mechanical waves. I anticipate my work to be a starting point to study the topological properties of mechanical waves. Even richer topological phases other than those have been found in electronics systems can be explored when we use more general mechanical waves e.g. waves in membranes with periodic densities. These phases may provide guides to hunt novel topological properties in other branches of science.
\end{abstract}

Both one dimensional(1D) and two dimensional(2D) systems have been used to simulate the topological insulators\cite{Haldane,Wang2009,Hafezi2013,Rechtsman,Mei,Khanikaev2013,Liang2013,
Albert,Ningyuan}. The edge states of 1D systems are zero dimensional and therefore the transport property of the edge states can't be studied in 1D simulations. However, this defect of 1D system is compensated by the fact that the parameter which is the analogue of the one component, say $k_y$, of the wave vector of the 2D system, can be readily controlled. In comparison, in 2D simulations, due to the lack of information of $k_y$, the transmission spectrums are measured instead\cite{Wang2009,Hafezi2013}. However, from the transmission spectrum we can't determine whether the edge states are topologically gapless or not.\\
Another difficulty of the electromagnetic simulations is that the topological invariant can't be directly measured. In electronic systems the topological invariant can manifest itself through the quantum Hall conductance\cite{Thouless1982}. However, due to the different statistics the quantum Hall conductance has no natural counterpart in  the electromagnetic waves. So the bulk topological invariant must be obtained from the Bloch waves. Although there are some theoretical and experimental works on how to measure the topological invariants in some indirect ways\cite{Bardyn2014,Hu2015,Mohammad2014,Alba2011,Dmitry2013}, no attempt has been made to obtain the bulk topological invariant by directly measuring the Bloch waves.\\
The above difficulties make the observation of the topological phases in electromagnetic systems not so conclusive. I find if classical strings are used to simulate topological phases, all these difficulties disappears. First, the eigenfrequencies of standing waves in the classical string can be easily measured and the parameter corresponding to $k_y$ can be readily controlled. So gaplessness of the edge states can be confirmed by directly measuring the spectrum of the edge states as function of $k_y$.\\
Second, the standing wave function with fixed boundary conditions can be easily measured. The other linearly independent solution of the eigenequation with same eigenvalue can be derived from it\cite{Simmons}. By proper linear combinations Bloch waves can be obtained from them. In this way the bulk topological invariants can be easily observed.\\
Due to their great controllability, two variable parameters can be used in strings to simulate topological phases in the three dimensional(3D) electronic systems. Such simulations have never been proposed in the electromagnetic systems.\\
Now I show the quantum Hall state can be simulated by a classical string with periodic density. The eigenvalue equation for the string can be  expressed as:
\begin{equation}\label{e1}
\frac{d^2}{dx^2}\phi(x)+\frac{\omega^2}{\rho(x,k_y)}\phi(x)=0
\end{equation}
$\rho(x,k_y)$ is a periodic function of $x$. $k_y$ is a parameter for 2D mapping similar to $\theta$ in Ref\cite{Mei} or $\phi$ in Ref\cite{Yaacov2012}. Here I assume $\rho(x)=m_0+m_1\cos(2\pi x/a-k_y)$. $a$ is the prime period of the density function. When $k_y$ adiabatically change by $2\pi$ the string is translated by one period. If the periodic boundary condition is used, the eigenfunctions of (1) are Bloch waves $\psi_{k_xk_y}(x)=e^{ik_xx}u_{k_xk_y}(x)$. The $k_x$ is a genuine wave vector. $k_x\in [0,2\pi/a]$ and $k_y\in[0,2\pi]$ form a 2D Brillouin zone. The eigenfrequencies as functions of $k_x$ and $k_y$ form 2D frequency bands as is shown in Fig.1 \textbf{a}.\\
The topology of a band can be described by it's Chern number\cite{Thouless1982,Thouless1983}
\begin{equation}\label{e2}
C=\frac{1}{2\pi}\int\int_{BZ} dk_x dk_y F(k_x,k_y)
\end{equation}
Where $F(k_x,k_y)$ is the Berry curvature
\begin{equation}\label{e3}
F(k_x,k_y)=i\int_0^adx(\frac{\partial u^{*}_{k_xk_y}}{\partial k_x}\frac{\partial u_{k_xk_y}}{\partial k_y}-\frac{\partial u^{*}_{k_xk_y}}{\partial k_y}\frac{\partial u_{k_xk_y}}{\partial k_x})
\end{equation}
$u_{k_xk_y}(x)$ satisfies $\int_0^a dx |u_{k_xk_y}(x)|^2=2\pi/a$. The Berry curvature for the lowest band is shown in Fig.1 \textbf{b}. By integrating the Berry curvature over the 2D  Brillouin zone one can find the Chern number of the lowest band is $1$. In this way the Chern number is obtained from the Bloch waves.\\
According to the bulk and edge correspondence\cite{Hatsugai1993,Hatsugai1993a} there must be gapless edge states in the bulk frequency gap. The gapless edge states spectrum as function of $k_y$ is plotted in Fig.2 \textbf{a} and \textbf{b}. Because the boundaries of the string can be easily controlled, even some of  delicate features of the edge state spectrum can be simulated. For example, when the length of the string isn't a multiple of $a$, a shift of the edge states similar to that in the incommensurate case in Ref\cite{Hatsugai1993a} appears as is shown in Fig.2 \textbf{b}.\\
Another interesting observation is that, similar to the discrete case, $x=na$ is always a zero point of the edge state wave function for all integer $n$. So as long as the length of the string equal a multiple of $a$ and the fixed boundary condition is used, the edge states eigen frequencies have nothing to do with the total length of string. This means the edge states spectrum of the string can be measured even when the string contains only one period, which facilitates the observation of the edge state spectrum greatly.\\
An adiabatic pumping is proposed in Ref\cite{Yaacov2012}. However, because the light isn't in an eigenstate of the Hamiltonian and the parameter $\phi$ doesn't change with time, the observation in Ref\cite{Yaacov2012} isn't an adiabatic pumping in it's original sense. To get more realistic adiabatic pumping we can start with an eigen edge state of the string and change $k_y$ slowly. Fig.2 \textbf{c} and \textbf{d} show that if the string is translated slow enough the edge state can be pumped from one end of the string to the other end and in this process the string is always approximately in an eigenstate of corresponding $k_y$.\\
In the electromagnetic simulations edge states is proved to be robust against the disorders\cite{Wang2009,Hafezi2013,Rechtsman}. However, the robustness of the gaplessness of the edge states hasn't been confirmed. I add some disorders to the density distribution of the string to simulate the 1D randomness in the 2D system in quantum mechanics as in Ref\cite{Hatsugai1993a}. As is shown in Fig.3 \textbf{a}, at least in this special condition, the edge states remain gapless when the disorders are present. As a by product I find some of the bulk states become localized when disorders are present as is shown in Fig.3 \textbf{b}. It's a perfect illustration of the Anderson localization\cite{Anderson}.\\
We can use two or more strings to simulate the quantum particles with spins. As an example, I use two strings to simulate the electrons. The two strings have same periodic density distributions except for opposite $k_y$ and they correspond to the two eigenstates the spin $S_z$. Thus the two strings simulate a $Z_2$ topological insulator with conserved $S_z$\cite{Kane2005a}.\\
By introducing coupling between the two strings the magnetic field can be simulated. We connect points with same $x$ by springs between the two strings. I assume the springs are distributed dense enough that they can be considered as a continuous distribution with spring rate $k(x)$. With these springs the eigenvalue equation can be expressed as:
\begin{equation}\label{e4}
    \left(
  \begin{array}{ccc}
          \rho(x,k_y)  & 0 \\
          0 & \rho(x,-k_y)
  \end{array}
  \right)\frac{d^2}{dx^2}
  \left(
  \begin{array}{c}
          \phi_1(x) \\
          \phi_2(x)
 \end{array}
 \right)
-
  \left(
  \begin{array}{ccc}
          \omega^2+k(x) & -k(x) \\
          -k(x) & \omega^2+k(x)
  \end{array}
  \right)
    \left(
  \begin{array}{c}
          \phi_1(x) \\
          \phi_2(x)
 \end{array}
 \right)=0
\end{equation}
 The coupling corresponds to $x$ dependent potential and magnetic field($x$ direction) in the 2D quantum system. When  $k(x)=k_0$ is a small constant a gap is opened in the edge states spectrum as is shown in the Fig.4 \textbf{a}, which confirms that only with the time reversal symmetry can the edge states of $Z_2$ topological insulator be truly gapless.\\
 A 3D weak topological insulator can be simulated if $k_0=\lambda \sin(k_z)$. Where $k_z\in[0,2\pi]$ and corresponds to the $z$ component of the wave vector. The coupling terms then become $\lambda \sin(k_z)I-\lambda \sin(k_z)\sigma_x$. Where $I$ is a $2\times2$ unit matrix and $\sigma_x$ is a Pauli matrix. Extra springs can be attached to the strings to obtain time reversal symmetric coupling terms $\lambda I-\lambda \sin(k_z)\sigma_x$. Then the strings are mapped to a time reversal symmetric 3D system. The edge state spectrum as function of $k_y$ and $k_z$ is plotted in Fig.4 \textbf{b}, which shows that there are two Dirac points at the symmetric point of the Brillouin zone. So the coupled strings are topologically equivalent to a weak topological insulator\cite{Fu}.\\
 Using more complicated density functions, a normal insulator, topological semimetal and weak topological insulator phase transition can be simulated. I assume the density function in (\ref{e4}) takes the form
 \begin{equation}\label{e5}
\rho(x,k_y)=m_0+[m_1(\cos(k_z)+1)+2m_2]\cos(2\pi x/a-k_y)+2m_3\cos(2\pi x/a-d)
\end{equation}
and the coupling term is still $\lambda I-\lambda \sin(k_z)\sigma_x$. In all the discussions $m_0$, $m_1$, $m_2$ and $m_3$ are positive numbers and $m_0>2(m_1+m_2+m_3)$. When $\lambda=0$, if $m_3>m_1+m_2$ the system corresponds a normal insulator. When $m_3=m_1+m_2$ the two bulk bands contact at $\textbf{k}=(\pi/a,\pi \pm d,0)$. If $m_1+m_2>m_3>m_2$ the system corresponds to a topological semimetal\cite{Wang,Liu,Wan,Burkov2011}. The semimetal has four Weyl points at $\textbf{k}=(\pi/a,\pi \pm d,\pm\arccos((2m_3-2m_2-m_1)/m_1))$ unless $d=0$ or $d=\pi$. When $d=0$ or $d=\pi$ the system has inversion symmetry and the Weyl points with same $k_z$ merge together and form Dirac points. When $m_3=m_2$ the Weyl or Dirac points with same $k_y$ merge together. The Weyl or Dirac points disappear if $m_3<m_2$ and system becomes a weak topological insulator. As long as $\lambda$ isn't too large the coupling terms don't affect the topological properties. They only relocate the Weyl points and split the Dirac points to Weyl points because they are inversion symmetry breaking. The Fermi arcs or Dirac cones of the surface state spectrum can all be observed in the strings with fixed boundary condition.\\
At last it may be pointed out my work not only provide a entirely new platform to study the topological phases in a convenient and accurate way, it may also find some engineering applications. For example, localized vibrations can be created and transferred by using a 1D periodic systems.

\begin{figure}[t]
\begin{center}
\ifthenelse{\boolean{SubmittedVersion}}{}{\includegraphics[width=120 mm]{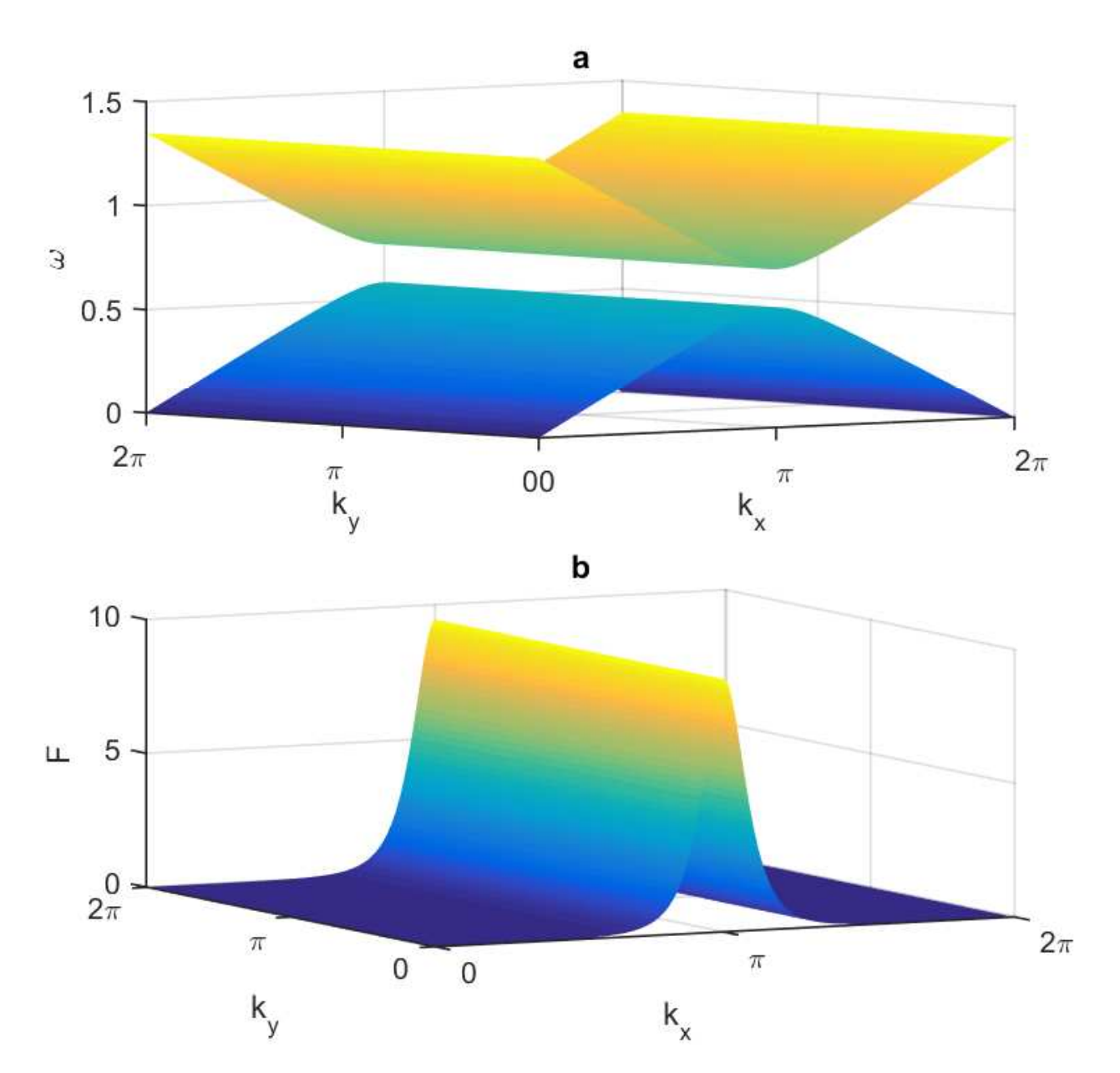}}
\end{center}
\caption{\textbf{Frequency bands and Berry curvature of lowest band.} \textbf{a},Lowest two eigen frequencies of string as functions of $k_x$ and $k_y$.\textbf{b},Berry curvature of the lowest band. In all the simulations in this paper $m_0=2$, $m_1=1$ and $a=2\pi$. } \label{Fig1}
\end{figure}

\begin{figure}[t!]
\begin{center}
\ifthenelse{\boolean{SubmittedVersion}}{}{\includegraphics[width=140 mm]{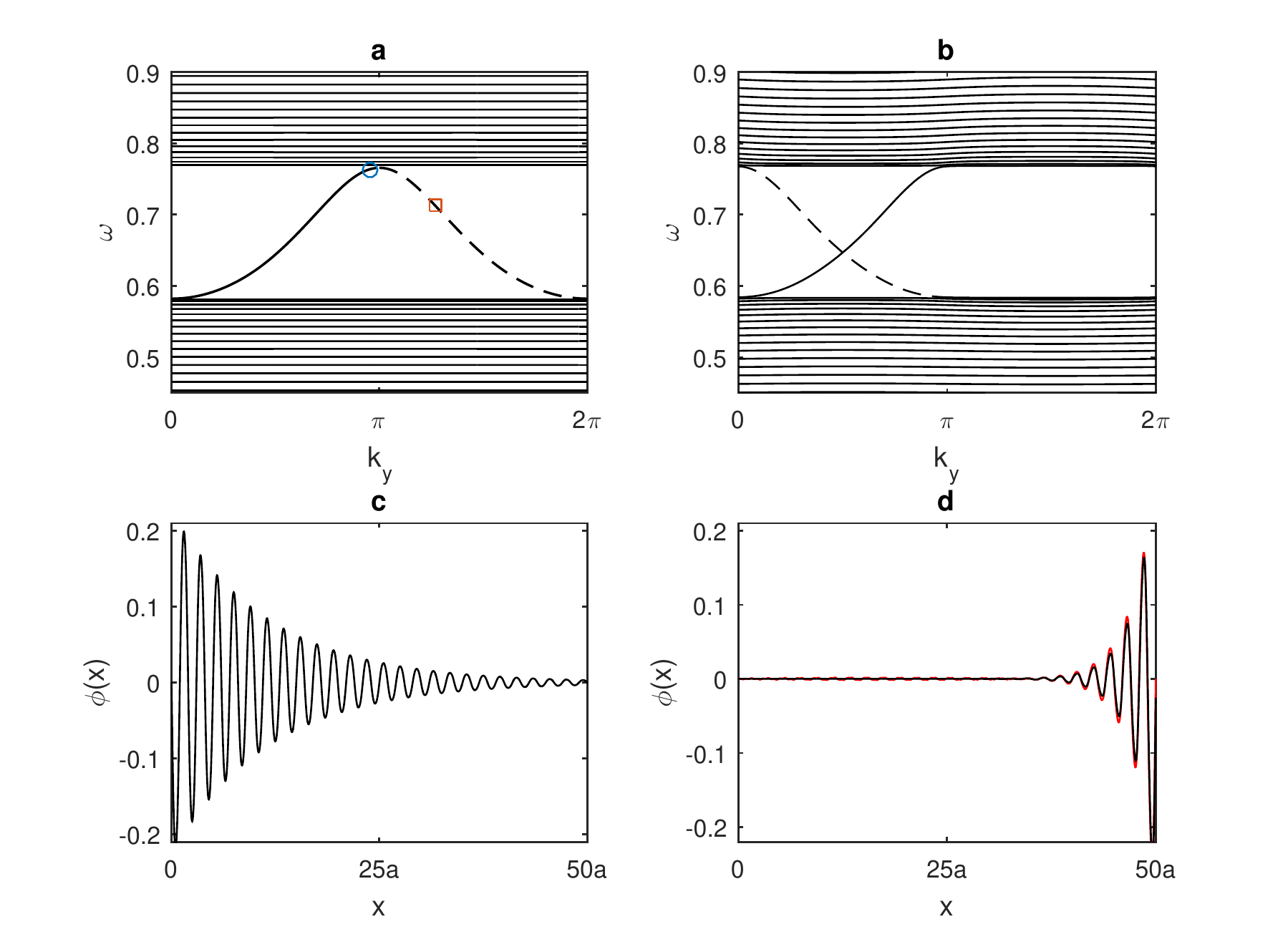}}
\end{center}
\caption{\textbf{Edge states spectrums and eigen wave functions of the edge states.} \textbf{a},\textbf{b} Eigen frequencies of (\ref{e1}) with fixed boundary condition. The solid(dashed) line in the gap denotes the edge states at left(right) end of the string. The length of the string $L=50a$ and $L=50.5a$ respectively in \textbf{a} and \textbf{b}. \textbf{c},\textbf{d} Edge state wave functions corresponding to the circle and square in \textbf{a} with $k_y=3$ and $k_y=4$. The red line in \textbf{d} depicts a wave function after $k_y$ slowly goes from $3$ to $4$. The initial wave function (with zero initial velocity) is the eigenfunction in \textbf{c} and the velocity of $k_y$ changing is $1.5\times10^{-5}$. This wave function is still approximately an edge state eigenfunction of (\ref{e1}) with $k_y=4$. }  \label{Fig2}
\end{figure}

\begin{figure}[t]
\begin{center}
\ifthenelse{\boolean{SubmittedVersion}}{}{\includegraphics[width= 140 mm]{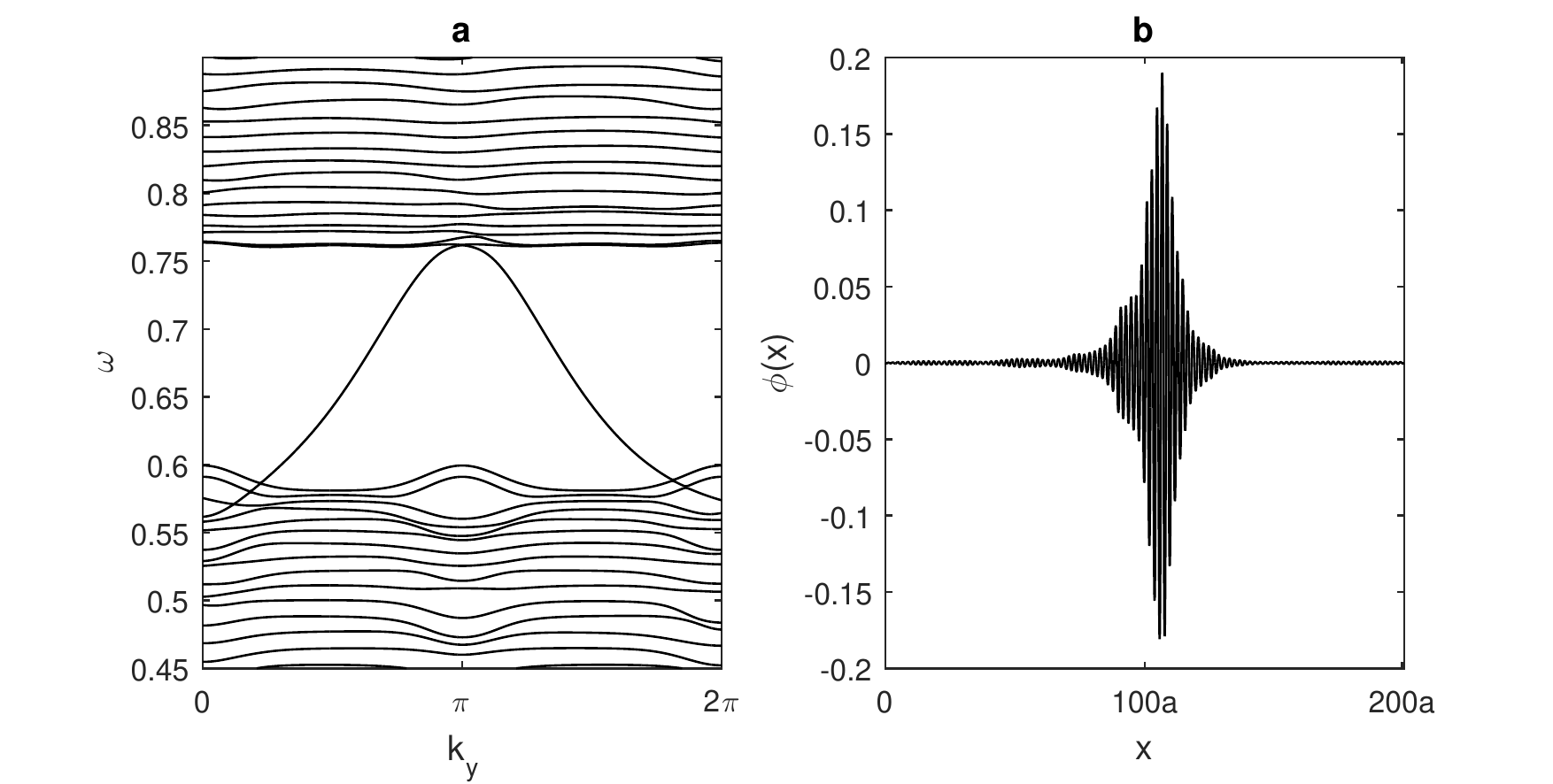}}
\end{center}
\caption{\textbf{Eigen frequencies and eigenfunctions of equation (\ref{e1}) with disorders added to the density distribution.} To simulated the disorders, I add random density distributions of type of $m_re^{-(x-x_n)^2/D^2}$ to $\rho(x,k_y)$. Where $m_r$ are random numbers uniformly distribute between $-0.5$ and $0.5$. $D=1$ and $x_n=(2n-1)\pi+k_y$ , $n$ take integers. \textbf{a} Eigen frequencies of (\ref{e1}) with disorders added to the density function. \textbf{b},A typical localized eigenfunction. }\label{Fig3}
\end{figure}

\begin{figure}[t]
\begin{center}
\ifthenelse{\boolean{SubmittedVersion}}{}{\includegraphics[width= 120 mm]{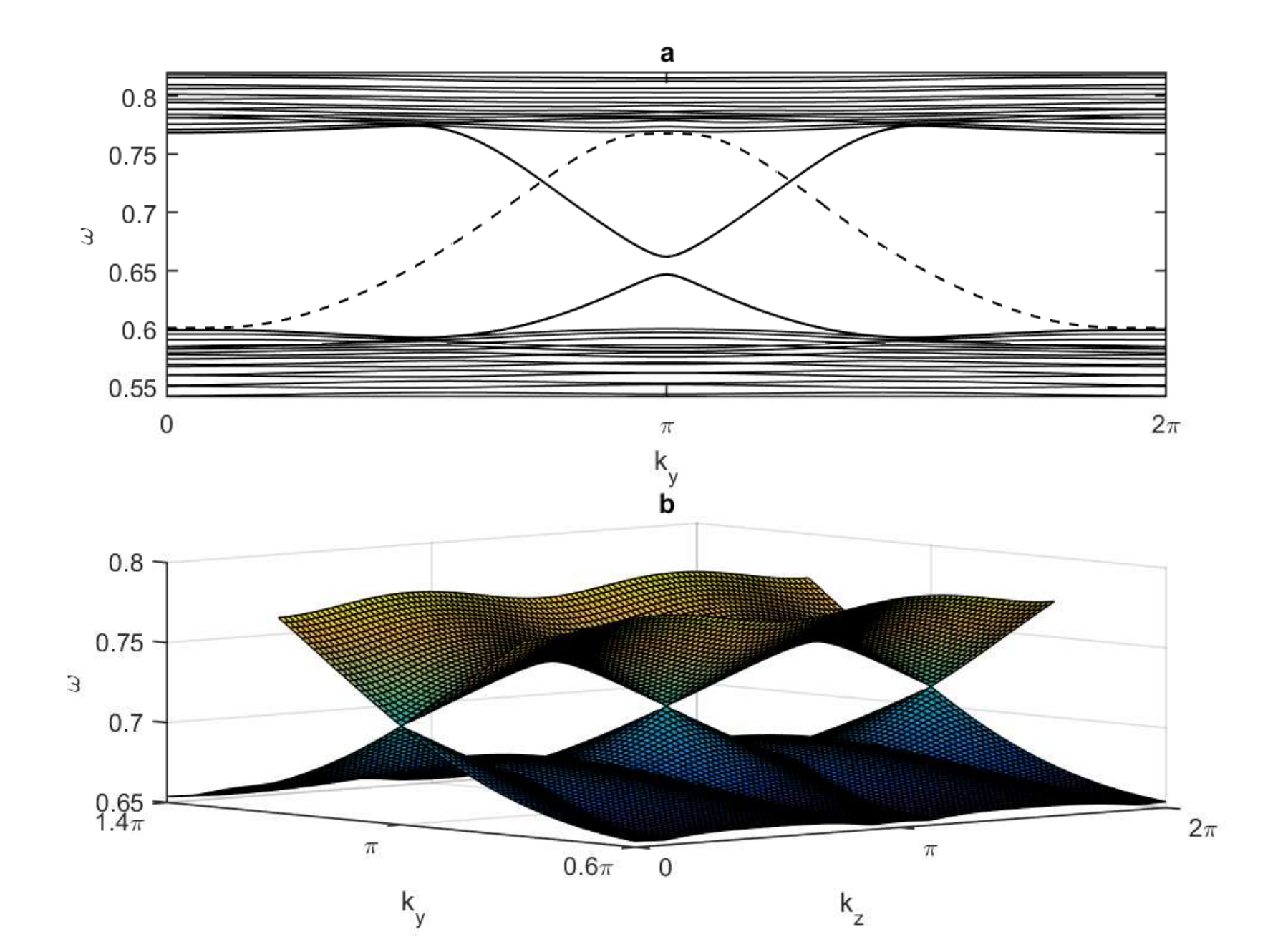}}
\end{center}
\caption{\textbf{Eigen frequencies of the coupled strings with fixed boundary condition.} \label{3 } \textbf{a},Edge state spectrum of coupled strings. $k_0=0.01$ and $L=50.25a$. The solid(dashed) line in the gap denotes the edge states at right(left) end of the string. The edge states at the left end of the string remain gapless because there is no coupling between states with different $k_y$.  \textbf{b},Edge states spectrum(right end) as function of $k_y$ and $k_z$. $\lambda=0.05$ and $L=50.25a$.  }\label{Fig3}
\end{figure}

\ifthenelse{\boolean{SubmittedVersion}}{\processdelayedfloats}{\cleardoublepage}

\end{document}